\documentstyle[11pt]{article}
\begin{document}
\large
\baselineskip=18pt

\thispagestyle{empty}
\vspace{.5cm}

\hfill DFTUSA 96/27

\begin{center}

\vspace{1cm}

{\LARGE \bf Symplectic Structures and Quantum Mechanics} 
\footnote[1]
{Supported in part by the italian Ministero
dell' Universit\`a e della Ricerca Scientifica e Tecnologica.\\
PACS Nos.:03.20+i,03.65-W.}
\end{center}
\medskip
\centerline{by}
\medskip

\begin{center}
{\bf Giuseppe Marmo} $^{1}$ and {\bf Gaetano Vilasi} $^{2}$
\end{center}
 
\bigskip
 
\begin{center}

$^{1}${\it Dipartimento di Scienze Fisiche , Universit\`a di Napoli,\\
Mostra d'Oltremare, Pad.19 - I-80125 Napoli, Italy.~(gimarmo@napoli.infn.it)}

\smallskip

$^{2}${\it Dipartimento di Fisica Teorica e smsa, 
Universit\`a di Salerno,\\
Via S. Allende, I-84081 Baronissi (SA), Italy.~(vilasi@vaxsa.csied.unisa.it)}
\par
and
\par
{\it Istituto Nazionale di Fisica Nucleare, Sezione di Napoli, Italy.}
\end{center}

\begin{abstract}
Canonical coordinates for the Schr\"odinger equation are introduced,
making more transparent its Hamiltonian structure. It is shown that 
the  Schr\"odinger equation, considered as a classical field theory, 
shares with  Liouville completely integrable field theories the 
existence of a {\sl recursion operator} which allows for the infinitely 
many conserved functionals pairwise commuting with respect to the 
corresponding Poisson bracket.
\par
The approach may provide a good starting point to get a clear 
interpretation of Quantum Mechanics in the general setting, provided 
by Stone-von Neumann theorem, of Symplectic Mechanics. 
It may give new tools to solve in the general case the inverse 
problem of quantum mechanics whose solution 
is given up to now only for one-dimensional systems by the 
Gel'fand-Levitan-Marchenko formula. 

\end{abstract}
 
\section{Introduction.}

In the past few years there has been a renewed interest in 
completely integrable Hamiltonian systems, specially in 
connection with the study of integrable quantum field theory, 
Yang-Baxter algebras and, more recently, quantum groups.
\par	
Loosely speaking,  completely integrable Hamiltonian systems 
are dynamical systems admitting a Hamiltonian description and 
possessing sufficiently many constants of motion so that they 
can be integrated by  quadratures.
\par	
For two-dimensional field theories, {\it a priori} criteria of 
integrability, have been established only by methods more 
directly related to group theory $^{1,2}$ and to familiar 
procedures of classical mechanics, looking at such systems as 
dynamics  on (infinite-dimensional) phase manifold $^{3,4,5,6,7,8}$
\par	
This point of view was also suggested by the occurrence in such 
models of a peculiar operator , the so called 
{\it recursion operator}  $^{9}$, relevant for the effectiveness 
of the method, which naturally fits in this geometrical setting as 
a mixed tensor field on the phase manifold M.
\par
In terms of such an operator the classical Liouville theorem on the 
integrability can be extended also to the infinite dimensional case. 
The same operator can be used to deal with Burgers equation $^{10}$.

\par
Some years ago it was suggested $^{11}$ the use of complex canonical 
coordinates in the formulation of a  generalized dynamics including
classical and quantum mechanics as special cases. In the same spirit
a somehow dual viewpoint is proposed: rather than to complexify 
classical mechanics it is useful to give a formulation of quantum 
mechanics in terms of {\sl realified} vector spaces.
\par
By using the Stone-von Neumann theorem a quantum mechanical system is 
associated with a vector field on some Hilbert space ({\sl Schr\"odinger
picture}) or a vector field, i.e. a derivation, on the algebra of 
observables ({\sl Heisemberg picture}).
\par
In classical mechanics the analog infinitesimal generator of 
canonical transformations is a vector field on a symplectic 
manifold (the {\sl phase space}).
\par
Therefore, if we want to use similar procedures, we need to real
off $L_2(Q, {\bf C})$, the Hilbert space  of square integrable
complex functions defined on the configuration space $Q$, 
as a symplectic manifold or, more specifically, as a cotangent bundle. 
We shall see that it can be considered as $T^*(L_2(Q, {\bf R}))$,
~~~~$L_2(Q, {\bf R})$ denoting the Hilbert space of square integrable
real functions defined on $Q$.
\par
This approach is different from previous ones $^{12}$ also 
dealing with the integrability of quantum mechanical system in
the Heisemberg and Schr\"odinger picture.
\par
In order to make more transparent the geometrical and the physical 
content of the paper difficult technical aspects, which are 
however important in the context of infinite dimensional manifold, 
as, for instance, the distinction $^{13}$ between {\sl weakly} 
and {\sl strongly} not degenerate bilinear forms, or the inverse 
of a Schr\"odinger operator and so on,  will  not be addressed. 
We shall limit ourselves to observe that no serious difficulties 
arise working on an infinite dimensional manifold whose local model 
is a Banach space, as in that case the {\sl implicit function 
theorem} still holds true.

\section{Complete Integrability and Recursion Operators} 

	Complete integrability of Hamiltonian systems with finitely many 
degrees of freedom is exhaustively characterized by the Liouville-Arnold 
theorem $^{14,15}$. An alternative characterization which may apply 
also to systems with infinitely many degrees of freedom can be 
given as follows.
Let M denote a smooth differentiable manifold, ${\cal X}(M)$ and 
$\Lambda(M)$ vector and covector fields on M.  
With any   $(1,1)$   tensor field   $T$   on  $ M$, two endomorphisms  
\par
${\hat T} : {\cal X}(M)\rightarrow {\cal X}(M)$~~~~~
and~~~~~~~ 
${\check T} : \Lambda(M)\rightarrow \Lambda(M) $
\par
\noindent
are associated:
\begin{equation}	
T(a, X)  = <\alpha,  {\hat T}X > = <{\check T}\alpha, X>,      
\end{equation}
with   $X$ and  $\alpha$ belonging to ${\cal X}(M)$  and  $\Lambda(M)$  
respectively.	The Nijenhuis  tensor  $^{16}$,  or torsion,  
of $T$ is  the (1,2) tensor field defined by:
\begin{equation}		
N_T(\alpha,X,Y) = <\alpha, H_T(X,Y)>                                       
\end{equation}
with the vector  field  $ H_T(X, Y)$   given by:
\begin{equation}		
H_T(X, Y)  = [\widehat{{\cal L}_{{\hat T}X}T} - {\hat T}
\widehat{{\cal L}_XT}]Y                                       
\end{equation}
${\cal L}_X$ denoting the Lie's derivative with respect to $X$.
\par
\noindent {\bf{ Integrability Criterium}}\footnote{The vector 
field $\Delta$ is not supposed to be Hamiltonian. Its 
Hamiltonian structure is generated by the hypothesis of the 
bidimensionality of the eigenspaces of T and $d \lambda \ne 0$.}
\smallskip
\par
{\it A dynamical vector field $\Delta$ which admits an invariant 
mixed tensor field T, with vanishing Nijenhuis tensor $N_{T}$ and 
bidimensional eigenspaces, completely separates in 1-degree 
of freedom dynamics. The ones associated with those degrees of 
freedom whose corresponding eigenvalues $\lambda$ are not stationary, 
are integrable and Hamiltonian $^{4}$.}
\par
An idea of the proof is given observing that the bidimensionality 
of eigenspaces
of $T$ and the condition $N_T=0$ imply the following form for $T$ 
\smallskip
\par
$T = \sum_{i} \lambda_{i} \biggl ({\delta\over \delta 
\lambda^{i}}\otimes \delta \lambda^{i} +{\delta\over \delta\phi^{i}} 
\otimes \delta \phi^{i} + {\delta\over \delta \phi^{i}} \otimes 
\delta \lambda^{i}\biggr ) + \sum^{2}_{\ell =1} 
\int_{0}^{k} dk~~ k {\delta\over \delta \psi^{\ell}_{k}}\otimes
\delta \psi^{\ell}(k)$
\smallskip
\par
The invariance of $T$ ~$ ({\cal L}_{\Delta} T = 0)$ implies for 
$\Delta$ the form
\par
$\Delta = \sum_{i=1}^{n} \Delta^{i}
(\lambda^{i}){\delta\over {\delta\phi^{i}}} + \sum^{2}_{\ell =1}  
\int dk \Delta^{\ell} (k)\biggl (\psi^{1}(k),\psi^{2}(k)\biggr )
{\delta\over {\delta\psi^{\ell}(k)}}$
\smallskip
\par
whose associated equations are:
\smallskip
\par
$\dot \psi^{1} (k) = \Delta^{1,k} (\psi^{1,(k)}, \psi^{2,(k)})$
\smallskip
\par
$\dot\psi^{2,(k)} = \Delta^{2,k}(\psi^{1,(k)},\psi^{2,(k)})$
\smallskip
\par
$\dot \phi^{i} = \Delta^{i} (\lambda^{i})$
\smallskip
\par
$\dot \lambda^{i}= 0$
\medskip
\par
For the discrete part of the spectrum of $T$  a 
symplectic form $\omega_{0}$ can be defined
$\omega_{0} = \sum_{i} f_{i} (\lambda^{i}) \delta \lambda^{i}
\wedge\delta\phi^{i}$  with respect to which the dynamics 
is a Hamiltonian one.
\medskip
\par
In next section the mentioned geometrical structures 
will be exhibited for the Schr\"odinger equation. 
\par

\section{Canonical Co\-or\-di\-nates for the Schr\"odin\-ger 
e\-qua\-tion}

Although in an infinite dimensional symplectic manifold a 
Darboux's chart, {\it a priori} does not exist, for the 
Schr\"odinger equation:

\begin{equation}
i\hbar{\partial \psi\over \partial t}=-{ {\hbar}^2\over 2m}
\triangle \psi + U({\bf r})\psi,
\end{equation}
natural 
canonical coordinates $p$ and $q$ can be introduced.
\par
We introduce the real and the imaginary part of the wave 
function $\psi$ :
\begin{eqnarray}
\cases{
p({\bf r},t) = {\sl Im}\psi({\bf r},t)\cr 
q({\bf r},t) = {\sl Re}\psi({\bf r},t)\cr}\nonumber,
\end{eqnarray}
and in this way $L_2(Q, {\bf C})$ is considered as the cotangent 
bundle of $L_2(Q, {\bf R})$.
\par
In these new coordinates, equation (4) takes the form:
\begin{equation}
{d\over dt}\pmatrix{p\cr q} = {1\over {\hbar}}\pmatrix{0&-1\cr 1& 0}
\pmatrix{{\delta H_1 \over\delta p}\cr {\delta H_1 \over \delta q}}
\end{equation}
where $H_1$ is defined by:
\begin{equation}
H_1[q,p]:={1\over 2}\int
d{\bf r}\{{{\hbar}^2\over m}[(\nabla p)^2 + (\nabla q)^2]+U({\bf r})
(p^2+q^2)\}
\end{equation}
and
${\delta H \over \delta q},~~{\delta H \over\delta p}$ denote
the components of the gradient of $H[q,p]$ with respect to the
real $L_2$ scalar product.
\par
Our system is then a Hamiltonian dynamical system with respect to 
the Poisson bracket defined for any two functionals $F[q,p]$ and 
$G[q,p]$ by:
\begin{equation}
\Lambda_1(\delta F, \delta G):=\{F,G\}_1:= {1\over {\hbar}}\int
d{\bf r}({\delta F \over \delta q}\cdot {\delta G \over \delta p}-
{\delta F \over \delta p}\cdot {\delta G \over \delta q})
\end{equation}
\par
What is less known is that the previous one is not the only possible 
Hamiltonian structure . 
As matter of fact the Schr\"odinger equation can also be written as:

\begin{equation}
{d\over dt}\pmatrix{p\cr q} = {1\over {\hbar}}\pmatrix{0&-{\cal H}\cr 
{\cal H}& 0}\pmatrix{{\delta H_0 \over\delta p}\cr {\delta H_0 \over 
\delta q}}
\end{equation}
where $H_0$ is defined by:

\begin{equation}
H_0[q,p]:={1\over 2}\int
d{\bf r}(p^2+q^2)
\end{equation}
and ${\cal H}$ is the Schr\"odinger operator:
\begin{equation}
{\cal H}:=-{{\hbar}^2\over 2m}\triangle + U({\bf r})
\end{equation}
\par
It is then again a Hamiltonian dynamical systems with a new Poisson bracket 
of any two functionals $F[q,p]$ and $G[q,p]$
given by:
\begin{equation}
\Lambda_0(\delta F, \delta G):=\{F,G\}_0:= \int
d{\bf r}({\delta F \over \delta q}\cdot {\cal H}{\delta G \over \delta p}-
{\delta F \over \delta p}\cdot {\cal H}{\delta G \over \delta q})
\end{equation}
So, with the same vector field, we have two choices:
\begin{itemize}
\item
A phase manifold with a universal symplectic structure:
\begin{equation}
\omega_1:= \hbar\int d{\bf r}(\delta p\wedge \delta q)
\end{equation}
and a Hamitonian functional depending on the classical potential.
\item
A phase manifold with a symplectic structure determined by the classical 
potential

\begin{equation}
\omega_0:= \hbar\int d{\bf r}({\cal H}^{-1}\delta p\wedge  \delta q)
\end{equation}
and the universal Hamiltonian functional representing the quantum 
probability.
\end{itemize}
The two brackets satisfy the Jacobi Identity, as the associated 2-forms 
are closed for they
do not depend on the point ($\psi\equiv (p,q)$) of the {\it phase space}.
\par
We have then the relation:
\begin{equation}
{\delta H_1 \over \delta u}= {\check T}{\delta H_0 \over \delta u}
\end{equation}
where:
\begin{equation}
 {\check T}:= \Lambda_1^{-1}\circ\Lambda_0=\pmatrix{{\cal H}&0\cr 0& {\cal H}}
\end{equation}
and
\begin{equation}
{\delta H \over \delta u}=
\pmatrix{{\delta H \over \delta q}\cr {\delta H \over \delta p}}
\end{equation}
As the tensor field T does not depend on the point ($\psi\equiv (p,q)$) 
of the {\it phase space}, its
torsion is identically zero, so that the relation (14) can be iterated to:
\begin{equation}
{\delta H_n \over \delta u}= {\check T}^n{\delta H_0 \over \delta u}
\end{equation}
It turns out that the Schr\"odinger equation admits infinitely many 
conserved functionals defined by:
\begin{equation}
H_n[q,p]:={1\over 2}\int 
d{\bf r}(p{\cal H}^np+q{\cal H}^nq)\equiv\int
d{\bf r}(\bar\psi{\cal H}^n\psi)
\end{equation}
They are all in involution with respect to the previous Poisson brackets:
\begin{equation}
\{H_n,H_m\}_0= \{H_n,H_m\}_1=0 
\end{equation}
This situation generalizes the one for finite dimensional Hamiltonian
systems $^{4}$.
It is worth to stress that for smooth potentials $U(x)$ in one space 
dimension, the eigenvalues of the Schr\"odinger operator ${\cal H}$  
are not degenerate and so the eigenvalues of $T$ are double degenerate.

\subsection{The eikonal transformation}
The transformation:

\begin{equation}
\cases{
p({\bf r},t) = A({\bf r},t)sinS({\bf r},t)\hbar^{-1}\cr
q({\bf r},t) = A({\bf r},t)cosS({\bf r},t)\hbar^{-1}\cr}
\end{equation}
is a canon\-i\-cal trans\-for\-ma\-tion be\-tween 
the $(p,q)$ co\-or\-di\-nates and $(\pi= S(2\hbar)^{-1}J, \chi=A^2)$, as:
\begin{equation}
\delta p\wedge \delta q = \delta ({S\over 2\hbar})\wedge \delta A^2 
\end{equation}

The Hamiltonian $H_1$ becomes:
\begin{equation}
K_1[\chi, \pi]=\int d{\bf r}\{{\hbar^2\over 2m}({(\nabla\chi)^2
\over 4\chi}+4\chi(\nabla\pi)^2)+U\chi\}
\end{equation}
and Hamilton's equations:

\begin{equation}
\left\{
\begin{array}{l}
{\partial \pi\over \partial t}=-{1\over {\hbar}}{\delta K_1 \over 
\delta \chi}\\{\partial \chi\over \partial t}=~{1\over {\hbar}}
{\delta K_1 \over\delta \pi}
\end{array}\right.,
\end{equation}
give: 

\begin{equation}
\left\{
\begin{array}{l}
{\partial \pi\over \partial t}={\hbar\over 2m}{\triangle(\sqrt\chi) 
\over \sqrt\chi}-
{\hbar\over m}(\nabla\pi)^2-U\hbar^{-1}\\
{\partial \chi\over \partial t}=-{2\hbar\over m}div(\chi\nabla\pi) 
\end{array}
\right.
\end{equation}
where  $P=\chi$ and ${\bf J}=\hbar\chi{\nabla S\over m}$ represent the 
{\it probability density} and the {\it current density} respectively.
\par
This transformation being nonlinear will transform previous biHamiltonian
descriptions into a mutually compatible pair of nonlinear type. They are
of $C$-type as introduced by Calogero $^{17}$.

\subsection{The quantum Lagrangians}

Having considered equations of motion for a quantum system as equations
for the integral curves of a vector field on a cotangent bundle, it is
a natural question to ask if this vector field may be associated with
a Lagrangian vector field on a tangent bundle. 
\par
This question for a Lagrangian Schr\"odinger Equation can be answered
as follows: 
\noindent
From equation (5) one gets Hamilton's equations:
\begin{equation}
\cases{
{\partial p\over \partial t} = -{1\over \hbar}{\cal H}q\cr
{\partial q\over \partial t} = ~~{1\over \hbar}{\cal H}p\cr}
\end{equation}
from which we derive the second order equation : 
\begin{equation}
{{\partial}^2 q \over \partial t^2}=-{1\over {\hbar}^2}{\cal H}^2q 
\end{equation}
The latter is the Euler-Lagrange equation associated with the Lagrangian 
functional:
\begin{equation}
L_1[q]={1\over 2}\int d{\bf r}dt(q_t^2-{1 \over \hbar^2}q{\cal H}^2q)
\end{equation}
Of course the Legendre transformation
\begin{equation}
\pi = {\delta L_1 \over \delta q_t}
\end{equation}
does not give the Hamilton's equation (25) but the related one:
\begin{equation}
\cases{
{\partial \pi \over \partial t} = -{1\over {\hbar}^2}{\cal H}^2q \cr
{\partial q\over \partial t} = \pi \cr}
\end{equation}
 Equations (25) follows straightforward from the Lagrangian $L_0$
given by:
\begin{equation}
L_0[q]={1\over 2}\int d{\bf r}dt(q_t{\cal H}^{-1}q_t-{1 \over \hbar^2}q
{\cal H}q)
\end{equation}

Of course $L_0$ is the Lagrangian which gives rise to the $\omega_0$ 
symplectic form and that:

\begin{equation}
{\delta L_1 \over \delta q}= {\cal H}{\delta L_0 \over \delta q}~;~~~~
~~~~~~~~~~~~~~~
{\delta L_1 \over \delta q_t}= {\cal H}{\delta L_0 \over \delta q_t}
\nonumber
\end{equation}
or equivalently:

\begin{equation}
{\delta L_1 \over \delta v}= {\check T}{\delta L_0 \over \delta v}
\end{equation}
where

\begin{equation}
{\delta L \over \delta v}:=\pmatrix{{\delta L \over \delta q}\cr 
{\delta L \over \delta q_t}}
\end{equation}

It is also clear that, as in the case of the Hamiltonian functionals,
relation (32) can be iterated to give altenative Lagrangian descriptions.

\section{Conclusions}

It has been shown as the Schr\"odinger equation, considered as a
vector field on an infinite dimensional vector space, admits more 
than one Hamiltonian formulation. Really it admits infinitely
many alternative Hamiltonian descriptions in terms of  
\begin{equation}
H_n[q,p]:={1\over 2}\int
d{\bf r}(p{\cal H}^np+q{\cal H}^nq)\equiv\int
d{\bf r}(\bar\psi{\cal H}^n\psi)
\end{equation}
and
\begin{equation}
\omega_n:= \hbar\int d{\bf r}({\cal H}^{n-1}\delta p\wedge  \delta q).
\end{equation}
providing us with the same vector field:
\begin{equation}
\Delta:= {1 \over \hbar}\int d{\bf r}({\cal H}p{\delta \over \delta q} -
{\cal H}q{\delta \over \delta p})
\end{equation}
defined by:
\begin{equation}
i_\Delta\omega_n:= -\delta H_n,
\end{equation}

These are associated with the Lagrangians

\begin{equation}
L_n[q]={1\over 2}\int d{\bf r}dt(q_t{\cal H}^{n-1}q_t-{1 \over \hbar^2}q
{\cal H}^{n+1}q)
\end{equation}
whose gradients are generated by the tensor field $T$.
\par
Even thought our construction is a formal one,
it is understood that the construction applies to any bounded, 
invertible operator ${\cal H}$.
\par
Finally, it is worth to stress that the Schr\"odinger equation, in  spite 
of its linearity, shows that the class of completely integrable field 
theories in higher dimensional spaces is not empty.
Moreover, previous analysis appears to be interesting also in the 
formulation of variational principles $^{18}$ for stochastic mechanics.

\par
\noindent 
{\Large \bf  Acknowledgement.}
\par
\medskip
The authors wish to thank prof. M.Rasetti for his encouragement and comments. 
One of them (G.V.) is in debt with prof. F.Guerra who stressed the 
relevance of the approach from the viewpoint of Stochastic Mechanics.
\newpage

\noindent {\Large\bf References}

\begin{description}      

\item{1.} 
R.Schmid, {\sl In\-fi\-nite Di\-men\-sional Hamil\-to\-nian Sys\-tems} 
(Bib\-liopo\-lis Naples 1987) and references therein. 

\item{2.}  
~JL.D.Faddeev and L.A.Takhtajan, {\sl Hamiltonian Methods in 
the Theory of Solitons}, (Berlin, Springer 1987), and references therein.

\item{3.}
~G. Vilasi, {\sl Phys. Lett.} {\bf B 94} (1980)195

G. Marmo, {\it A Geometrical Characterization of Completely 
Integrable Systems} in Ge\-om\-e\-try and Physics (Mo\-dugno ed.) 
(Pi\-ta\-go\-ra, Florence 1982)257.

S. De Filippo, G. Vilasi, {\it Geometrical Methods for Infinite 
Dimensional Dynamical Systems}, Proceedings  Second World Conference 
on Mathematics, (Las Palmas, Spain 1982)236.

S. De Filippo, G. Marmo, G. Vilasi, {\sl Phys. Lett.} {\bf B 117} 
(1982)418.

G. Marmo, {\it Nijenhuis Operators in Classical Dynamics}, 
Proceedings of Seminar on Group Theoretical Methods in Physics, 
USSR (1985). 

P. Di Stasio and G.Vilasi, {\sl Lett. in Math. Phys.} {\bf 11}(1986)299

G. Vilasi, {\sl Phys. Lett.} {\bf B 174} (1986)203

G. Marmo, G. Vilasi, {\sl Phys. Lett.} {\bf B 277} 
(1992)137.

\item{4.}
~S. De Filippo, G. Marmo, M. Salerno, G. Vilasi, 
{\sl On the Phase Manifold Geometry  of Integrable Nonlinear 
Field Theories}.IFUSA n.2 (1982)

S. De Filippo, G. Marmo, M. Salerno, G. Vilasi,
{\sl Il N. Cimento} {\bf 83 B}, 2 (1984)97.

S. De Filippo, M. Salerno, G. Vilasi, {\sl Lett. Math. Phys.} 
{\bf 9} (1985) 85.

G.Landi, G.Marmo and G.Vilasi, {\sl J.Math. Phys.} {\bf 35}(1994)808

\item{5.}
~I.M. Gel' fand and I. Ya Dorfman, {\sl Funct. Anal.} {\bf 14}, 3 (1980)71.
 
\item{6.} 
~F.Magri, {\sl J.Math. Phys}. {\bf 18}, (1978)1156; {\sl Lect. Notes in Math.} 
{\bf 120} (Berlin, Springer 1980)

\item{7.} 
~B. Fuchsteiner, {\sl Progr.\ of Theor. Phys.} {\bf 68}, 4 (1982)1082.

\item{8.}  
~V.E.Zakharov, B.G.Konopolchenko, {\sl Comm.Math.Phys.} {\bf 94} (1984)483
       
Y.Kosmann-Schwarzbach, {\sl Geom\`etrie des systems bihamiltonian}, Pub.IRMA, 
2,1   Lille(1986)

\item{9.}
~P.D.Lax, {\sl Comm.Pure Appl. Math.} {\bf 21} (1968)467;{\bf 28} (1975)141; 
{\sl Siam Rev.}, {\bf 18}(1976)351

\item{10.}
S.De Filippo, G.Marmo, M.Salerno, G.Vilasi,
{\sl Lettere a Il N.Cimento} {\bf 37}, 3 (1983)105. 

\item{11.} 
~F. Strocchi, {\sl Rev.Mod.Phys.} (1966)

\item{12.} 
~B.A. Kupershmidt, {\sl Phys. Lett.} {\bf A 109} (1985)136.

A.M. Bloch,{\sl Phys. Lett.} {\bf A 116} 
(1986)353.

R.Cirelli, L.Pizzocchero, {\sl On the integrability of quantum
mechanics as an infinite dimensional Hamiltonian system},
Nonlinearity {\bf 3} (1990)

G.Cassinelli, E.De Vito, A.Levrero, {\sl The dynamical evolution
of quantum systems in the adiabatic approximation}, Preprint 
Universit\`a di Genova (1996)

\item{13.} 
~R. Chernoff and J.Marsden, {\sl Lect. Notes in Math.}{\bf 425} (Berlin, 
Springer 1974)

\item{14.}
~J. Liouville, {\sl Acta Math.} {\bf 20} (1897)239.  

H. Poincar\`e, {\sl Acta Math.} {\bf 13} (1899)1.

\item{15.}
V.I.Arnold, {\sl Les m\`ethodes math\`ematiques de la M\`ecanique Classique} 
(Mir, Moscou 1976)

R.Abraham and J.E.Marsden, {\sl Foundations of Mechanics} (Benjamin, 
Massachussets 1978)

\item{16.}
~A.Frolicher and A.Nijenhuis, {\sl Indag. Math.}, {\bf 23},(1956)338

A.Nijenhuis, {\sl Indag. Math.}, {\bf 49},(1987)2

\item{17.}
~F.Calogero, {\sl Physica} {\bf 18D},(1986)280

\item{18.}
~F.Guerra and L.Morato, {\sl Phys.Rev.D}{\bf 27}(1983)1774

F.Guerra and R.Marra, {\sl Phys.Rev.}{\bf D28} (1983)1916; {\bf D29} (1984)1647

\end{description}

\end{document}